\documentstyle[editedvolume,numreferences,psfig]{crckapb} 

\newcommand{\be}{\begin{equation}}
\newcommand{\ee}{\end{equation}}
\newcommand{\bea}{\begin{eqnarray}}
\newcommand{\eea}{\end{eqnarray}}
\newcommand{\les}{\ell_{\hbox{\tiny ES}}}

\newcommand{\pnuc}{p_{\hbox{\tiny nuc}}}
\newcommand{\peff}{p_{\hbox{\tiny A}}^{\hbox{\tiny eff}}}
\newcommand{\pth}{p_{\hbox{\tiny S}}}
\newcommand{\Pdep}{P_{\hbox{\tiny dep}}}
\newcommand{\pA}{p_{\hbox{\tiny A}}}
\newcommand{\pB}{p_{\hbox{\tiny B}}}

\newcommand{\omf}{\omega_{\hbox{\tiny MF}}}
\newcommand{\ttr}{\tau_{tr}}
\newcommand{\tdep}{\tau_{dep}}
\newcommand{\tres}{\tau_{res}}
\newcommand{\Nall}{N_{\hbox{\tiny all}}}
\newcommand{\Ndis}{N_{\hbox{\tiny dis}}}
\newcommand{\Wni}{W_{\hbox{\tiny NI}}}
\newcommand{\fra}[2]{\hbox{${#1\over #2}$}}

\begin{opening}
\title{IRREVERSIBLE NUCLEATION IN MULTILAYER GROWTH}

\author{PAOLO POLITI}
\institute{INFM, UdR Firenze\\
           L.go E. Fermi 2\\ 
           I-50125 Firenze, Italy}

\author{CLAUDIO CASTELLANO}
\institute{INFM, UdR Roma ``La Sapienza"\\
           P.le A. Moro 2\\
           I-00185 Roma, Italy}

\end{opening}

\runningtitle{IRREVERSIBLE NUCLEATION}

\begin{document}


\begin{abstract}
The epitaxial growth process of a high symmetry surface
occurs because adatoms meet and nucleate new islands, that
eventually coalesce and complete atomic layers.
During multilayer growth, nucleation usually takes place
on top of terraces where the geometry of the diffusion
process is well defined:
We have studied in detail the spatiotemporal distribution
of nucleation events and the resulting nucleation rate,
a quantity of primary importance to model experimental
results and evaluate diffusion barriers at step-edges.
We provide rigorous results for irreversible nucleation
and we assess the limits of mean-field theory (MFT):
we show that MFT overestimates the correct result by a factor
proportional to the number of times an adatom diffusing on the
terrace visits an already visited lattice site.
In this report we aim at giving a simple physical account of our results.
\end{abstract}

\section{Introduction}

Crystal growth by Molecular Beam Epitaxy can be schematically
depicted~\cite{libroMBE,libroJV} as a process of uniform
deposition of particles,
their surface diffusion, and their incorporation
at the growing surface. Incorporation means attachment to another
diffusing adatom or to a growing island or to a preexisting
step: The two first cases are typical of a high symmetry surface
where steps are continuously created and destroyed by the growth process,
while attachment to a preexisting train of steps is the growth mode
of a vicinal surface.

Here we investigate the process of attachment
between diffusing adatoms, called `nucleation' because the so formed dimer
may be the nucleus of a new terrace~\cite{Venables,Brune}. 
On a high symmetry surface 
nucleation may be considered as the first step of the growth process,
followed by the capture of other adatoms and by the coalescence of
neighbouring islands. This is surely true for the very first stages
of growth (the so called submonolayer regime) where it is possible 
to separate the three mentioned steps by monitoring the adatom and 
island densities~\cite{AFL}.
In this regime nucleation takes place on the flat substrate and 
nucleation events are not independent processes:
The formation of a nucleus reduces the adatom density in its
surroundings and therefore disfavors further nucleations.
In a sense, nucleation sites repel each other.

Beyond the submonolayer regime, nucleation mainly occurs on top terraces: 
in this case, therefore, the diffusion process and the possible nucleus 
formation take place in a confined region of well defined geometry.
This makes the theoretical study of the nucleation process
on top of a terrace easier than on a flat surface.
However, the spatio-temporal distribution of nucleation
events cannot directly be related to experiments; in order to obtain
experimentally relevant information it is necessary to complement the
results for a top terrace with the growth dynamics of such a terrace.
Because of this the full study of `terrace nucleation' is not a
`single terrace' problem (we will come back to this issue in the next
Section).

We have not mentioned so far the possibility that adatom attachment
may not be the end of the story. 
In fact, depending on the substrate
temperature, the intensity of the incoming flux and the surface symmetry,
dimer formation may be an irreversible process or not. 
Irreversible nucleation means that once two adatoms meet
they stop diffusing and do not detach: 
in the following we are going to study this case.
Furthermore islands will be supposed to be compact:
this requires that attachment of
adatoms to islands is followed by step-edge diffusion allowing the
search of high coordination sites along the edge of the 
terrace~\cite{PRL_Evans}.

A short report on the results that will be presented here has already
appeared~\cite{corto} and a paper with all the mathematical details
of the calculations is being prepared~\cite{lungo}.
Here we aim at giving a simple physical description of our approach, of the
results and of the open questions.
We will consider mainly two quantities: the nucleation rate $\omega$ and the
spatial distribution $P(n)$ of nucleation events. 
$\omega$ is defined as the number of nucleation events per unit time 
on the whole terrace (it is therefore proportional to the area $L^d$ 
of the island); $P(n)$ is a {\it normalized} quantity ($\sum_n P(n)=1$)
that tells how nucleations are spatially distributed on the sites $n$
of the terrace\footnote{In $d=2$ the index $n$ should be meant as a
pair of integer indices $(n_x,n_y)$.}.
A third quantity is of interest as well~\cite{corto}:
the probability $Q(t)$ that nucleation takes place at time $t$
after the deposition of the second adatom.
It has less experimental relevance than $\omega$ and $P(n)$ and therefore
it will not be discussed here.

\section{Time scales and basic assumptions}
\label{sec:2}

The process of nucleation on a top terrace involves three typical
time scales, whose
detailed discussion can be found in Ref.~\cite{KPM}.
If $F$ is the intensity of the incoming flux, $D$ the adatom surface
diffusion constant, $L$ the linear size of the terrace and $d=1,2$ the
dimension of the terrace, we define $\tdep=(FL^d)^{-1}$ as the 
average time interval between deposition events on the terrace,
$\ttr\sim L^2/D$ as the typical time taken by the adatom
to travel through the terrace,
and $\tres$ as the typical time an adatom stays on the terrace before
getting off. This last quantity depends on the boundary conditions at the
terrace edge, i.e. on the possible existence of an Ehrlich-Schwoebel (ES) 
barrier pushing back adatoms~\cite{KE}; by introducing the 
interlayer transport rate $D'$ we can define the so called
ES length $\les=({D\over D'} -1)$~\cite{libroJV} and in terms of it 
the regimes of weak and strong
barriers correspond to $\les\ll L$ and $\les\gg L$, respectively.
The time $\tres$ has to do with the average stationary adatom density 
$\bar\rho$ on the island~\cite{KPM}: $\tres=\bar\rho/F\simeq
L(L+\alpha_d \les)/D$ where $\alpha_d$ is a numerical factor
dipending on the dimension and the shape of the terrace. 
It is straightforward that $\tres\sim\ttr$ for absorbing boundaries
($\les=0$).

The relation $\ttr\ll\tdep$ is always verified in realistic growth conditions
and therefore, according to the value of the ratio $\les/L$ we can
distinguish three different regimes:
({\it i})~$\ttr\sim\tres\ll\tdep$ (weak ES effect);
({\it ii})~$\ttr\ll\tres\ll\tdep$ (strong ES effect);
({\it iii})~$\ttr\ll\tdep\ll\tres$ (infinite ES effect).  

The word `infinite' for the third regime is to be
intended in physical terms: it means that if an atom is on the 
terrace a second atom will surely arrive before the first one
leaves the terrace, so that the nucleation rate is just the inverse
of $\tdep$.

Since we study {\it irreversible} nucleation and 
$\ttr\ll\tdep$, we can limit ourselves to study nucleation
as a ``two adatoms" process disregarding 
processes involving three adatoms or more. 
We can show the validity of this assumption in the most unfavorable
case of infinite ES barriers: a first atom is
deposited at time $t=0$ and stays on the terrace;
after an average time $\tdep$ a second atom comes: they
meet in a typical time $\ttr$ and the probability that a third atom
arrives in the meanwhile is negligible just because 
$\ttr/\tdep\ll 1$. 

A further issue deserves to be discussed before the illustration of
our approach.
We study quantities concerning a terrace of {\it fixed} size $L$:
it is obvious that a given island grows in time, but the nucleation rate
$\omega(L)$ is evaluated keeping $L$ constant.
The expressions for $\omega$ and $P(n)$ are general
and do not depend on the details of the growth process that determine
the actual time dependence of the terrace size $L(t)$.
These details do enter in the problem if ---for example--- we want to
compute the probability $f(t)$
that a nucleation event has occurred before time $t$~\cite{TDT}, because the
evaluation of $f(t)=1-\exp\{-\int_0^t d\tau \omega(L(\tau))\}$
requires the knowledge of $L(\tau)$.

\section{The nucleation rate: method and results}

We have argued that it is sufficient to study two adatoms processes.
This means that the following picture applies: Adatoms arrive on the
terrace at a rate $FL^d=\tdep^{-1}$ and stay there an average time
$\tres$; a nucleation event takes place if an adatom is still on the
terrace when the next one is coming {\it and} they meet before
getting out. It is therefore possible to define a nucleation
probability per atom $\pnuc$ such that $\omega=\tdep^{-1}\cdot\pnuc$.
If $\Pdep(t)=\tdep^{-1}\exp(-t/\tdep)$ is the probability that a
second atom is deposited a time $t$ after a first one, we have
\be
\pnuc = \int_0^\infty dt \Pdep(t)\tilde\pnuc(t)~,
\ee
where $\tilde\pnuc(t)$ is the probability that the first
atom $A$ deposited at time zero and the second atom $B$ deposited
a time $t$ later meet.

In the same manner, if we are looking for the spatial probability 
distribution $P(n)$ of nucleation sites we need to evaluate the
integral
\be
P(n) = \int_0^\infty dt \Pdep(t)\tilde P(n;t)~,
\ee
where $\tilde P(n;t)$ is the distribution evaluated for the first
atom $A$ deposited at time zero and the second atom $B$ deposited
a time $t$ later. 

Both $\tilde\pnuc(t)$ and $\tilde P(n;t)$ depend {\it linearly}
on the initial `probability distributions' for atoms $A$ and $B$.
Therefore we only need to evaluate $\tilde\pnuc(t)$ and $\tilde P(n;t)$
for two particles deposited simultaneously with an effective initial 
distribution~\cite{corto} for particle $A$:
\be
\peff(n) = \int_0^\infty dt \Pdep(t) \pA(n,t)~,
\label{eq:peff}
\ee
where $\pA(n,t)$ is the distribution of a particle evolving {\em alone}
on the terrace for time $t$:
it is the dynamical evolution of the single
particle probability $\pA(n,0)=1/L^d$ via the discrete diffusion equation:
\be
\pA(n,t+1) = \fra{1}{2d} \sum_\delta \pA(n+\delta,t)~,
\ee
where ``$n+\delta$" labels the neighboring sites of ``$n$".

We can explain the meaning of Eq.~(\ref{eq:peff}) easily:
atom $A$ is deposited uniformly at time zero; atom $B$ has a probability
$\Pdep(t)$ to be deposited at time $t$ and therefore it has probability
$\Pdep(t)$ to find atom $A$ on the terrace with distribution $\pA(n,t)$.
It is worth stressing that $\pA(n,t)$ is not normalized:
its sum on all sites, $S(t)$, is the probability that atom $A$ is still
on the terrace when $B$ comes in.
In conclusion, we have to study the problem of two atoms diffusing on the
terrace, whose starting distributions are $\peff(n)$ and $\pB(n)=1/L^d$,
respectively.

The effective distribution carries two pieces of information:
its integral (${\cal I}=\sum_n\peff(n)$) determines the scaling of the
nucleation rate $\omega$ while its normalized version ($\pth(n)$)
influences the shape of $P(n)$.

Using the definition (\ref{eq:peff}) and the explicit knowledge
of $\Pdep(t)$ we can write down a differential equation for $\peff(n)$,
whose solution is possible~\cite{lungo} for any $\les$.
We propose here a simpler argument: The normalization factor of $\peff(n)$
is given by ${\cal I}=\tdep^{-1}\int_0^\infty dt \exp(-t/\tdep) S(t)$ 
where $S(t)$ is exactly the probability that atom $A$ is still on the
terrace at time $t$. For weak ES barriers the exponential can be taken
as a constant on the time scale of the decay of $S(t)$ and
${\cal I}=\tdep^{-1}\int_0^\infty dt S(t) = \tres/\tdep$.
For strong ES barriers $S(t)$ decays exponentially~\cite{Harris} 
and ${\cal I}= \tres/(\tres+\tdep)$. The solution of the differential
equation mentioned above confirms~\cite{lungo} 
this result for any $\les$.

It is therefore possible to write $\peff(n)=\fra{\tres}{\tres+\tdep}
\pth(n)$: The question is now how $\pth(n)$ looks like. In the limit
of weak ES barriers we can repeat the above argument and find that
$\peff(n)=\tdep^{-1}\int_0^\infty \pA(n,t)$. The time integral of
the single particle probability distribution is nothing but the
solution of the stationary diffusion equation in the presence of a
constant flux~\cite{corto}, which is known to have a parabolic
form~\cite{SH}.
We can conclude that for any $\les$:
\be
\peff(n)= {\tres \over \tres+\tdep} \pth(n)~,
\ee
where $\pth(n)$ is the normalized steady state distribution.
Consequently the nucleation probability per atom takes the form:
\be
\pnuc = {\tres \over \tres+\tdep} W
\label{pnuc_ex}
\ee
where $W$ is the probability that two adatoms $A$ and $B$, deposited
simultaneously with distributions $\pA(n)=\pth(n)$ and $\pB(n)=1/L^d$,
meet before descending.
$W$ is almost independent from the exact spatial profiles
of atoms $A$ and $B$ and it mainly depends on $\les/L$ and on the 
space dimensionality $d$.

Let us now assume $\tres\ll\tdep$, i. e. consider weak and strong
ES barriers (for the `infinite ES barrier' regime see below).
Then,
\be
\pnuc \simeq {\tres \over \tdep} W \simeq \bar\rho L^d W
\simeq \bar\rho\Ndis~,
\label{pnuc_app}
\ee
where we have used the relation $\bar\rho=F\tres$ and the 
relation~\cite{lungo}
$W\simeq\Ndis/L^d$ between the probability $W$ that two adatoms meet and
the number $\Ndis$ of {\it distinct} sites visited by an adatom
during its diffusion on the terrace. 

The relation $\pnuc\simeq \bar\rho\Ndis$ can be intuitively
justified~\cite{JPsolo} with the following argument:
the nucleation probability per atom is given by the number
of distinct sites ($\Ndis$) visited by each atom times the probability
that a given site is occupied ($\bar\rho$).
This argument breaks down if the average number of adatoms present at the
same time on the terrace (and equal to $\bar\rho L^d$) is larger than
one; such condition is equivalent to $\tres\gg\tdep$
(regime ({\it iii})).

The nucleation rate is finally written as:
\be
\omega(L) \simeq {\tres W\over \tdep^2 } \simeq FL^d\bar\rho\Ndis
\label{omega_ex}
\ee
and we want to compare it with the well known mean field 
result~\cite{Venables,TDT}:
\be
\omf = D L^d \bar\rho^2 = FL^d\bar\rho\Nall~,
\label{omega_cm}
\ee
where we have made use of the relation~\cite{JPsolo} 
$\bar\rho=\fra{F}{D}\Nall$,
$\Nall$ being the {\it total} number of sites visited by an adatom
during its diffusion on the terrace.
The comparison of Eqs.~(\ref{omega_ex}) and (\ref{omega_cm}) is fully
transparent: $\omf/\omega \simeq \Nall/\Ndis\equiv{\cal N}$, 
that is mean field
theory overestimates the correct nucleation rate by a factor proportional
to the number of times an adatom diffusing on the island
visits an {\it already} visited site.

Our comparison gets complete once we introduce a model where
diffusing adatoms do {\it not} interact: even if they meet each adatom
keeps diffusing until they get off. If the average number of fictitious
nucleations between non interacting adatoms is $\Wni$ 
we simply have $\Nall\simeq\Wni L^d$ and
$\omf=(\tres \Wni/ \tdep^2)$, i.e. mean field theory ---as expected
by the relation $\omf = D L^d \bar\rho^2$--- treats adatoms as independently
diffusing particles.

The above results are valid for $\tres\ll\tdep$ because they derive 
from Eq.~(\ref{pnuc_app}) rather than from the more general
Eq.~(\ref{pnuc_ex}). What does it happen in the regime of `infinite'
ES effect?
In this case $W=1$ and $\pnuc=1$ so that
$\omega = \tdep^{-1} = FL^d = F \Ndis$
while $\omf=D L^d \bar\rho^2 \simeq F L^d \bar\rho \Nall$.
Hence $\omf/\omega\simeq \bar\rho L^d \Nall/\Ndis$. 
We can sum up our results for the nucleation rate by reporting the
ratio $\omf/\omega$ for the three regimes introduced in
Sec.~\ref{sec:2}, both in $d=1$ and $d=2$:
this is done in Table~1 where
we use the already mentioned relation between $\Nall$ and $\bar\rho$,
the well known results~\cite{Hughes} that in absence of ES barriers
$\Ndis\simeq L$ in $d=1$ and $\Ndis\simeq L^2/\ln L$ in $d=2$ and finally
that $\Ndis=L^d$ for $\les\gg L$.

\begin{table}[htb]
\begin{center}
\caption{Ratio $\omf/\omega$ for the three relevant regimes.
For ({\it i}) and ({\it ii}), $\omf/\omega={\cal N}\equiv\Nall/\Ndis$;
for ({\it iii}) $\omf/\omega\simeq\bar\rho L^d {\cal N}$ where
$\bar\rho\simeq\fra{F}{D}L\les$ is the average adatom density.}
\begin{tabular}{cccc}
\hline
 & ({\it i}) & ({\it ii}) & ({\it iii}) \\
 & $~\tres\simeq\ttr ~$ & $~~\ttr\ll\tres\ll\tdep ~~$ & $~\tres\gg\tdep ~$ \\
\hline
$d=1$ & $L$     & $\les$ & $F (L \les)^2/D$  \\
$d=2$ & $\ln L$ & $\les/L$ & $F (L \les)^2/D$ \\
\hline
\end{tabular}
\end{center}
\end{table}

The result $\omf/\omega=\les/L$ ---valid in the two dimensional strong
barrier regime--- had already been found in Ref.~\cite{KPM}.
In that paper logarithmic corrections were neglected, which corresponds to
disregarding the $L$-dependence of the factor $W$ in the weak barriers regime.

\section{The spatial distribution of nucleation events}

In this Section we study what sites are the most
favored for nucleation. The result is strongly dependent on the 
exact spatial
profiles of the initial probability distributions 
for particles $A$ and $B$ ($p_{\hbox{\tiny A,B}}(n)$):
we have justified below Eq.~(\ref{eq:peff}) that $A$ is distributed
as the stationary solution of the diffusion equation
($\pA(n)=\pth(n)$) while $B$ is uniformly distributed
($\pB(n)=1/L^d$).

The dynamical evolution of two diffusing particles on a 
$d-$di\-men\-sio\-nal terrace can be easily mapped on the problem
of a single walker in a space of dimension $d'=2d$.
If $m$ and $n$ label the positions of the two atoms a nucleation event
occurs when $m=n$, i.e. when the ``single walker" crosses the diagonal
of a square terrace if $d=1$ ($d'=2$) or the `diagonal plane' of a four
dimensional hypercube if $d=2$ ($d'=4$).

In $d=1$ we have solved analytically~\cite{corto} 
the problem for the limit cases of zero and infinite ES effect; 
for any value of $\les$ we can compute exact numerical results
for $P(n)$ both in $d=1$ and $d=2$~\cite{lungo}.
Our main results in $d=2$
are reproduced in Fig.~1 where we give the resulting spatial
distribution of nucleation events along the diagonal of a square
terrace and we compare it with Mean Field Theory:
$P(n)\sim\pth^2(n)$.
In absence of ES barriers (Fig.~1a) MFT works remarkably well
but its accuracy gets worse with increasing $\les$ (Figs.~1b,c)
and MFT fails completely for large ES barriers (Fig.~1d):
in this case MFT predicts that $P(n)$ gets flat while a clearly
rounded shape is obtained.

Fig.~1 confirms that MFT is equivalent to an ``independent adatoms"
model and this is the reason of its failure with increasing $\les$:
for $\les\gg L$ independent adatoms perform many more fictitious
nucleations than for $\les\ll L$ (see Table~1).

\begin{figure}
\centerline{\psfig{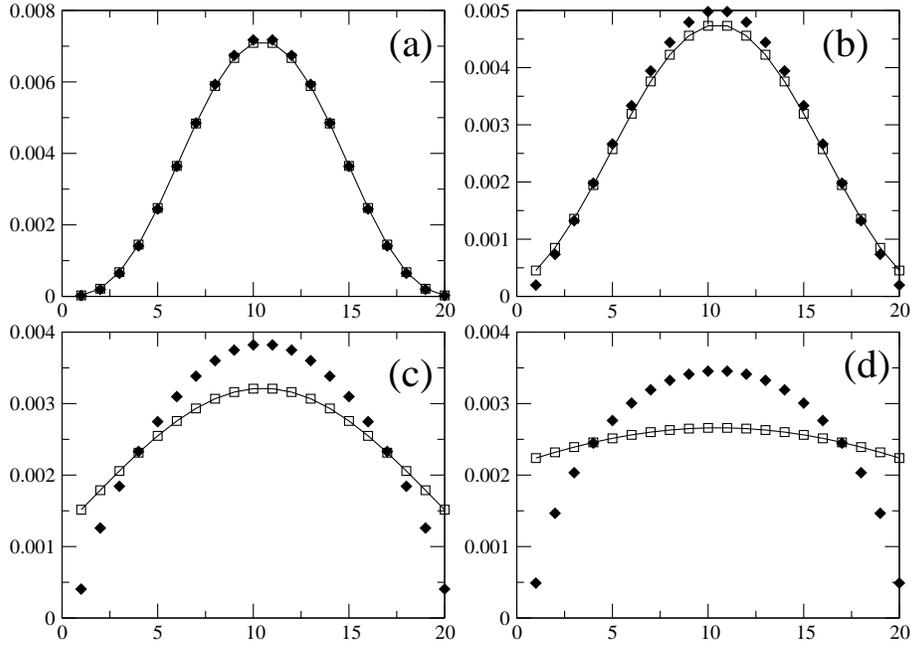}} 
\caption{$P(n)$ along the diagonal of a square terrace of size $L=20$
for (a)~$\les=0$, (b)~$\les=4$, (c)~$\les=20$, (d)~$\les=100$.
Full diamonds: exact theory; open squares: mean field theory;
full line: ``independent adatoms" model.
}
\end{figure}

\section{Conclusions}

We think that two main aspects of our work should be emphasized.
Limits and inaccuracies of mean field theory have been pointed out
by several authors during the years~\cite{PVW,PRL_Wolf,KPM,Maass,Krug}
but we have performed a rigorous study for any value of the ES
barriers in one and two dimensions.
Furthermore we study the spatial distribution $P(n)$ of nucleation
events and we are not aware of previous similar analyses.
The second aspect is that we provide a simple physical interpretation
of the failure of mean field theory: it counts all the nucleations of two
independently diffusing adatoms.

The nucleation rate is of primary experimental relevance because most
of the methods to determine step-edge barriers ---or to evaluate other 
quantities derived from $\les$--- require the knowledge of
$\omega(L)$~\cite{TDT,altro}.
It would be therefore useful to reconsider such derivations which made
use of $\omf$: this has already been started in Ref.~\cite{KPM}.

As for $P(n)$, a direct experimental determination is extremely 
complicated~\cite{Michely}.
Nonetheless its knowledge has a theoretical interest because $P(n)$
may enter in mesoscopic models of crystal growth~\cite{EVPV,Ratsch},
i.e. in models where surface diffusion is not taken into account
microscopically but through a mesoscopic surface current plus the
rule for nucleating new terraces.

Finally we mention a few extensions of our work that are presently
in progress: Firstly, the problem of nucleation 
between adatoms of different species,
i.e. particles having different diffusion constants. 
Secondly, the study of $P(n)$ in the limit $L\to\infty$ in order to
understand what features are maintained in this limit. 
Finally, the nucleation on top of a fractal terrace.

\end{document}